\documentclass[12pt,preprint]{aastex}
\shorttitle{Birthrates and delay times of SNe Ia} \shortauthors{WANG
et al.}

\begin{document}
\title{Birthrates and delay times of Type Ia supernovae}

\author{B. Wang\altaffilmark{1,2,3}, Z. Liu\altaffilmark{1,2,3}, Y. Han\altaffilmark{1,2,3}, Z. Lei\altaffilmark{1,2,3}, Y. Luo\altaffilmark{1,2,3} and Z. Han\altaffilmark{1,2}}
\altaffiltext{1}{National Astronomical Observatories/Yunnan
Observatory, the Chinese Academy of Sciences, Kunming 650011, China;
wangbo@ynao.ac.cn} \altaffiltext{2}{Key Laboratory for the Structure
and Evolution of Celestial Objects, the Chinese Academy of Sciences,
Kunming 650011, China} \altaffiltext{3}{Graduate University of the
Chinese Academy of Sciences, Beijing 100049, China}

\begin{abstract}
Type Ia supernovae (SNe Ia) play an important role in diverse areas
of astrophysics, from the chemical evolution of galaxies to
observational cosmology. However, the nature of the progenitors of
SNe Ia is still unclear. In this paper, according to a detailed
binary population synthesis study, we obtained SN Ia birthrates and
delay times from different progenitor models, and compared them with
observations. We find that the Galactic SN Ia birthrate from the
double-degenerate (DD) model is close to those inferred from
observations, while the birthrate from the single-degenerate (SD)
model accounts for only about 1/2$-$2/3 of the observations. If a
single starburst is assumed, the distribution of the delay times of
SNe Ia from the SD model is a weak bimodality, where the WD + He
channel contributes to the SNe Ia with delay times shorter than
$100$\,Myr, and the WD + MS and WD + RG channels to those with age
longer than 1\,Gyr.
\end{abstract}


\section{Introduction}
Type Ia supernovae (SNe Ia) appear to be good cosmological distance
indicators due to their high luminosities and remarkable uniformity,
and thus are used for determining cosmological parameters(e.g.,
Riess et al. 1998; Perlmutter et al. 1999). They also throw light on
the understanding of galactic chemical evolution owing to the main
contribution of iron to their host galaxies. However, several key
issues related to the nature of their progenitor systems and the
physics of the explosion mechanisms are still not well
understood(Hillebrandt \& Niemeyer 2000; Wang et al. 2008a), and no
SN Ia progenitor system before the explosion has been conclusively
identified. This may raise doubts about the distance calibration,
which is purely empirical and based on the SN Ia sample of the low
red-shift universe.

It is generally believed that SNe Ia are thermonuclear explosions of
carbon--oxygen white dwarfs (CO WDs) in binaries. Over the past few
decades, two families of SN Ia progenitor models have been proposed,
i.e., the double-degenerate (DD) and single-degenerate (SD) models.
The DD model involves the merger of two CO WDs, where the total mass
of the two CO WDs is larger than the Chandrasekhar (Ch) mass limit
(Webbink 1984; Han 1998). For the SD model, the companion is
probably a main-sequence (MS) star or a slightly evolved subgiant
star (WD + MS channel), or a red giant star (WD + RG channel), or an
He star (WD + He star channel) (Hachisu et al. 1999; Li $\&$ van den
Heuvel 1997; Han $\&$ Podsiadlowski 2004; Chen $\&$ Li 2009; Meng et
al. 2009; L\"{u} et al. 2009; Wang et al. 2009a; Wang, Li $\&$ Han
2010). Note that, recent observations have suggested that at least
some SNe Ia can be produced by a variety of progenitor systems
(e.g., Patat et al. 2007; Wang et al. 2008b).

At present, various progenitor models of SNe Ia can be examined by
comparing the distribution of the delay time (between the star
formation and SN Ia explosion) expected from a progenitor model with
that of observations (e.g., Chen $\&$ Li 2007; Wang \& Han 2010).
There are three important observational results for SNe Ia, i.e.,
the strong enhancement of the SN Ia birthrate in radio-loud
early-type galaxies, the strong dependence of the SN Ia birthrate on
the colors of the host galaxies, and the evolution of the SN Ia
birthrate with redshift. Mannucci et al. (2006) found that the
observational results can be best matched by a bimodal delay time
distribution, in which about half of the SNe Ia explode soon after
starburst, with a delay time less than $\sim$100\,Myr, while those
remaining have a much wider distribution with a delay time of
$\sim$3\,Gyr. It is suggested that 10\% (weak bimodality) to 50\%
(strong bimodality) of all SNe Ia belong to the young SNe Ia
(Mannucci et al. 2008). The existence of the young and old
populations of SNe Ia is also supported by some other observations
(e.g., Aubourg et al. 2008; Totani et al. 2008).

The purpose of this paper is to study SN Ia birthrates and delay
times from different progenitor models, and compare them with
observations. We describe the binary population synthesis (BPS)
approach for the different progenitor models in Section 2 and
present the BPS results in Section 3. Section 4 ends the paper with
a discussion.

\section{Binary population synthesis}
In order to investigate SN Ia birthrates and delay times, we
performed a series of Monte Carlo simulations in the BPS study. In
each simulation, we followed the binary evolution via the rapid
binary evolution code developed by Hurley et al. (2002).

\subsection{Common envelope in binary evolution}
The progenitor of an SN Ia is a close WD binary, which has most
likely emerged from the common envelope (CE) evolution of a giant
binary. During the binary evolution, the primordial mass ratio
(primary to secondary) is crucial for the mass transfer. If it is
larger than a critical mass ratio, $q_{\rm c}$, the mass transfer
may be dynamically unstable and a CE develops. The mass ratio
$q_{\rm c}$ varies with the evolutionary state of the primary at the
onset of the Roche lobe overflow (RLOF) (e.g., Han et al. 2002). In
this study we adopt $q_{\rm c}=4.0$  when the primary is in the
main-sequence stage or Hertzsprung gap. This value is supported by
detailed binary evolution studies (e.g., Han et al. 2000). If the
primary is on the first giant branch or asymptotic giant branch
stage, we use
\begin{equation}
q_{\rm c}=[1.67-x+2(\frac{M_{\rm c1}^{\rm P}}{M_{\rm 1}^{\rm
P}})^{\rm 5}]/2.13,
  \end{equation}
where $M_{\rm 1}^{\rm P}$ is the mass of the primary, $M_{\rm
c1}^{\rm P}$ is the core mass of the primary, and $x={\rm d}\ln
R_{\rm 1}^{\rm P}/{\rm d}\ln M_{\rm 1}^{\rm p}$ is the mass-radius
exponent of the primordial primary and varies with composition. If
the mass donor stars (primaries) are naked He giants, $q_{\rm
c}=0.748$  based on eq. (1) (see Hurley et al. 2002 for details).

When a CE forms, embedded in the CE is a `new' binary consisting of
the dense core of the primary and the secondary. Owing to frictional
drag within the envelope, the orbit of the `new' binary decays and a
large part of the orbital energy released in the spiral-in process
is injected into the envelope. The CE ejection is still an open
problem. Here, we use the standard energy equations (Webbink 1984)
to calculate the output of the CE phase. The CE is ejected if
\begin{equation}
 \alpha_{\rm ce} \left( {G M_{\rm don}^{\rm f} M_{\rm acc} \over 2 a_{\rm f}}
- {G M_{\rm don}^{\rm i} M_{\rm acc} \over 2 a_{\rm i}} \right) = {G
M_{\rm don}^{\rm i} M_{\rm env} \over \lambda R_{\rm don}},
\end{equation}
where $\lambda$ is a structure parameter that depends on the
evolutionary stage of the mass donor star, $M_{\rm don}$ is the mass
of the donor, $M_{\rm acc}$ is the mass of the accretor, $a$ is the
orbital separation, $M_{\rm env}$ is the mass of the donor's
envelope, $R_{\rm don}$ is the radius of the donor, and the indices
${\rm i}$ and ${\rm f}$ denote the initial and final values. The
right side of the equation represents the binding energy of the CE,
while the left side shows the difference between the final and
initial orbital energy, and $\alpha_{\rm ce}$ is the CE ejection
efficiency (the fraction of the released orbital energy used to
eject the CE). For this prescription of the CE ejection, there are
two highly uncertain parameters (i.e., $\lambda$ and $\alpha_{\rm
ce}$). As in previous studies, we combine $\alpha_{\rm ce}$ and
$\lambda$ into one free parameter $\alpha_{\rm ce}\lambda$ (Wang et
al. 2009b).

\subsection{The progenitor model of SNe Ia}
Theoretically, there is a consensus that an SN Ia is the explosion
and complete disintegration of a CO WD that has a mass close to the
Chandrasekhar limit mass $\rm M_{Ch}$ (1.378\,$M_{\odot}$ in this
work). We adopt two popular models for the progenitors of SNe Ia:
the DD and SD models.

\subsubsection{The DD model}
For the DD model, SNe Ia arise from the merging of two close CO WDs
that have a combined mass larger than or equal to the $\rm M_{\rm
Ch}$ (Webbink 1984; Han 1998). Both WDs are brought together by
gravitational wave (GW) radiation on a timescale $t_{\rm GW}$
(Landau \& Lifshitz 1971),
\begin{equation}
t_{\rm GW}(\rm yr \it)=\rm 8\times10^{7}\it(\rm yr \it
)\times\frac{(M_{\rm 1}+M_{\rm 2})^{\rm 1/3}}{M_{\rm 1}M_{\rm
2}}P^{\rm 8/3}(\rm h),
\end{equation}
where $P$ is the orbital period in hours, $t_{\rm GW}$ in years and
$M_{\rm 1}$, $M_{\rm 2}$ in $M_{\odot}$.

The delay time from the star formation to the occurrence of an SN Ia
is equal to the sum of the timescale on which the secondary star
becomes a WD and the orbital decay time $t_{\rm GW}$. Here, we set
the $\alpha_{\rm ce}\lambda=1.5$, which reproduces the number of the
double-degenerate objects in the Galaxy (Hurley et al. 2002).

\subsubsection{The SD model}
For the SD model, we considered the WD + He star, WD + MS and WD +
RG channels in this paper. Take the WD + He star channel for
example, we followed the evolution of $1\times10^{\rm 7}$ sample
binaries from the star formation to the formation of the WD + He
star systems. We assumed that, if the parameters of a CO WD + He
star system at the onset of the RLOF are located in the SN Ia
production regions in the ($\log P^{\rm i}, M^{\rm i}_2$) plane
(Figure 8 of Wang et al. 2009a), where $P^{\rm i}$ and $M^{\rm i}_2$
are the orbital period and the mass of the He companion star at the
onset of the RLOF, respectively, an SN Ia is produced. Note that,
the method of the BPS study for the WD + MS and WD + RG channels is
similar to that of the WD + He star channel. For the SN Ia
production regions of these two channels, see Wang, Li \& Han
(2010). Here, we set the $\alpha_{\rm ce}\lambda=0.5$, which is our
standard model for the formation of these SD channels (Wang, Li \&
Han 2010).

\subsection{Basic parameters for Monte Carlo simulations}
In the BPS study, the Monte Carlo simulation requires as input the
initial mass function (IMF) of the primary, the initial mass-ratio
distribution, the distribution of initial orbital separations, the
eccentricity distribution of binary orbit, and the star formation
rate (SFR) (Wang et al. 2009b; Han 2008a,b).

(1) The IMF of Miller \& Scalo (1979) is adopted. The primordial
primary is generated according to the formula of Eggleton et al.
(1989),
\begin{equation}
M_{\rm 1}^{\rm p}=\frac{0.19X}{(1-X)^{\rm 0.75}+0.032(1-X)^{\rm
0.25}}M_{\rm \odot},
  \end{equation}
where $X$ is a random number uniformly distributed in the range [0,
1] and $M_{\rm 1}^{\rm p}$ is the mass of the primordial primary,
ranging from 0.1\,$M_{\rm \odot}$ to 100\,$M_{\rm \odot}$.

(2) The initial mass-ratio distribution of the binaries, $q'$, is
quite uncertain for binary evolution. For simplicity, we take a
constant mass-ratio distribution,
\begin{equation}
n(q')=1, \hspace{2.cm} 0<q'\leq1,
\end{equation}
where $q'=M_{\rm 2}^{\rm p}/M_{\rm 1}^{\rm p}$.

(3) We assume that all stars are members of binaries and that the
distribution of separations is constant in $\log a$ for wide
binaries, where $a$ is separation and falls off smoothly at a small
separation
\begin{equation}
a\cdot n(a)=\left\{
 \begin{array}{lc}
 \alpha_{\rm sep}(a/a_{\rm 0})^{\rm m}, & a\leq a_{\rm 0},\\
\alpha_{\rm sep}, & a_{\rm 0}<a<a_{\rm 1},\\
\end{array}\right.
\end{equation}
where $\alpha_{\rm sep}\approx0.07$, $a_{\rm 0}=10\,R_{\odot}$,
$a_{\rm 1}=5.75\times 10^{\rm 6}\,R_{\odot}=0.13\,{\rm pc}$ and
$m\approx1.2$. This distribution implies that the numbers of wide
binaries per logarithmic interval are equal, and that about 50\,per
cent of stellar systems have orbital periods less than 100\,yr (Han
et al. 1995).

(4) A circular orbit is assumed for all binaries. The orbits of
semidetached binaries are generally circularized by the tidal force
on a timescale which is much smaller than the nuclear timescale.
Also, a binary is expected to become circularized during the RLOF.

(5) We simply assume a constant SFR over the past 15\,Gyr, or,
alternatively, a delta function, i.e., a single starburst. In the
case of the constant SFR, we calibrate the SFR by assuming that one
binary with a primary more massive than $0.8\,M_{\odot}$ is formed
annually (see Han et al. 1995; Hurley et al. 2002). From this
calibration, we can get ${\rm SFR}=5\,M_{\rm \odot}{\rm yr}^{-1}$.
For the case of the single starburst, we assume a burst producing
$10^{11}\,M_{\odot}$ in stars. In fact, a galaxy may have a
complicated star formation history. We choose only these two
extremes for simplicity. A constant SFR is similar to the situation
of spiral galaxies (e.g., Han \& Podsiadlowski 2004), while a delta
function to that of elliptical galaxies or globular clusters.

\section{The results of binary population synthesis}
\subsection{The birthrates of SNe Ia}

We performed four sets of simulations with metallicity $Z=0.02$ and
${\rm SFR}=5\,M_{\rm \odot}{\rm yr}^{-1}$ to systematically
investigate Galactic birthrates of SNe Ia for the DD and SD models,
where the SD model includes the WD + He star, WD + MS and WD + RG
channels (see Figure 1). We find the birthrate from the DD model
$\sim$$2.9\times 10^{-3}\ {\rm yr}^{-1}$, only slightly lower than
the birthrate inferred from observations (i.e., 3$-$$4\times
10^{-3}\ {\rm yr}^{-1}$; Cappellaro \& Turatto 1997), while the
total birthrates from the SD models can only account for about
1/2$-$2/3 of the observations, where the birthrate from the WD + He
star channel $\sim$$0.3\times 10^{-3}\ {\rm yr}^{-1}$, the WD + MS
channel $\sim$$1.8\times 10^{-3}\ {\rm yr}^{-1}$ and the WD + RG
channel $\sim$$3\times 10^{-5}\ {\rm yr}^{-1}$.

The SN Ia birthrate in galaxies is the convolution of the
distribution of the delay times (DDT) with the star formation
history (SFH):
\begin{equation}
\nu(t)=\int^t_0 SFR(t-t')DDT(t')dt',
\end{equation}
where the $SFR$ is the star formation rate, and $t'$ is the delay
times of SNe Ia. Due to a constant SFR adopted in this paper, the SN
Ia birthrate $\nu(t)$ is only related to the $DDT$, which can be
expressed by
\begin{equation}
DDT(t)=\left\{
\begin{array}{lc}
0, & t<{\rm t_1},\\
DDT'(t) , &   {\rm t_1} \leq t \leq{\rm t_2},\\
0, & t>{\rm t_2},\\
\end{array}\right.
\end{equation}
where ${\rm t_1}$ and ${\rm t_2}$ are the minimum and maximum delay
times of SNe Ia, respectively, and the $DDT'$ is the distribution of
the delay times between ${\rm t_1}$ and ${\rm t_2}$. If $t$ is
larger than ${\rm t_2}$, eq. (7) can be written as
\begin{equation}
\nu(t)={\rm SFR}\int^{\rm t_2}_{\rm t_1}DDT'(t')dt'={\rm constant}.
\end{equation}
Therefore, the SN Ia birthrates shown in Figure 1 seem to be
completely flat after the first rise.

\subsection{The delay times of SNe Ia}

Figure 2 displays the evolution of SN Ia birthrates for a single
starburst with a total mass of $10^{11}\,M_{\odot}$. In the figure
we see that SNe Ia from the DD model have the delay times of
$\sim$89\,Myr$-$15\,Gyr, which are close to the observational
results from Totani et al. (2008). However, it is suggested that the
DD model is likely to lead to an accretion-induced collapse rather
than to an SN Ia (Nomoto \& Iben 1985). Thus, the DD model is not
supported theoretically. We also find that SNe Ia from the SD models
have a wide distribution of the delay times, where the WD + He star
channel contributes to the SNe Ia with delay times shorter than
$100$\,Myr, and the WD + MS and WD + RG channels to those with age
longer than 1\,Gyr (the WD + MS channel also contributes to the SNe
Ia with intermediate delay times $\sim$100\,Myr$-$1\,Gyr). Note
that, Chen \& Li (2007) studied the WD + MS channel by considering a
circumbinary disk which extracts the orbital angular momentum from
the binary through tidal torques. This study also provides a
possible way to produce such old SNe Ia ($\sim$1$-$3\,Gyr).

The SD model is currently a favorable progenitor model of SNe Ia.
The distribution of the delay time from the SD models is similar to
that derived from observations by Mannucci et al. (2006), except
that the peak value of young population is smaller than that in
Mannucci et al. (2006). The WD + He star channel produces 14 per
cent of all SNe Ia, which constitutes the weak bimodality as
suggested by Mannucci et al. (2008). The results from the SD models
seem to be slightly smaller than those in Totani et al. (2008).
However, given the high possibility of errors in the observation by
Totani et al. (2008), our results are close to observations.

\section{Discussion}

In our BPS studies, we assumed that all stars are in binaries and
about 50\,per cent of stellar systems have orbital periods less than
100\,yr. In fact, this is known to be a simplification. The binary
fractions may depend on metallicity, environment and spectral type.
If we adopt 40\,per cent of stellar systems with orbital periods
below 100\,yr by adjusting the parameters in eq. (6), we estimate
that the Galactic SN Ia birthrate from the DD model will decrease to
be $\sim$$2.3\times 10^{-3}\ {\rm yr}^{-1}$, the WD + He star
channel $\sim$$0.24\times 10^{-3}\ {\rm yr}^{-1}$, the WD + MS
channel $\sim$$1.4\times 10^{-3}\ {\rm yr}^{-1}$ and the WD + RG
channel $\sim$$2.4\times 10^{-5}\ {\rm yr}^{-1}$.

Hachisu et al. (2008) investigated new evolutionary models for SN Ia
progenitors, introducing the mass-stripping effect on a MS or a
slightly evolved companion star by winds from a mass-accreting WD.
Though the model offers a possible way to produce young SNe Ia, the
model depends on the efficiency of the mass-stripping effect. We
also find that the model produces very few young SNe Ia by a
detailed BPS approach. Thus, we consider the WD + He star channel as
a main contribution to the young population of  SNe Ia, possible the
whole young population.

The Galactic SN Ia birthrate from the WD + RG channel is
$\sim$3$\times 10^{-5}\ {\rm yr}^{-1}$, which is low compared with
observations. Thus, SNe Ia from this channel may be rare. However,
further studies on this channel are necessary, since this channel
may explain some SNe Ia with long delay times. RS Oph and T CrB,
both recurrent novae are probable SN Ia progenitors and belong to
the WD + RG channel (e.g. Belczy$\acute{\rm n}$ski \& Mikolajewska
1998; Hachisu et al. 1999, 2007). Detecting Na I absorption lines
with low expansion velocities indicates that the companion of the
progenitor of SN 2006X may be an early RG star (Patat et al. 2007).
We note that a symbiotic star with aspherical stellar wind may also
provide a way for the evolution of the WD + RG systems towards SNe
Ia (L\"{u} et al. 2009).

Provided that the DD model can produce SNe Ia, an explosion
following the merger of two WDs would leave no remnant, while the
companion star in the SD model would survive and potentially be
identifiable (Wang \& Han 2009). Thus, it will be a promising method
to test different SD models of SNe Ia by identifying their surviving
companions.

The young and old populations of SNe Ia may have an effect on models
of galactic chemical evolution, since they would return large
amounts of iron to the interstellar medium much earlier or later
than previously thought. In future investigations, we will explore
the detailed influence of SNe Ia on the chemical evolution of
stellar populations.

\acknowledgments This work is supported by the National Natural
Science Foundation of China (Grant No. 10821061), the National Basic
Research Program of China (Grant No. 2007CB815406), and the Yunnan
Natural Science Foundation (Grant No. 08YJ041001).

\clearpage

\begin{figure*}
\includegraphics[width=8.5cm,angle=270]{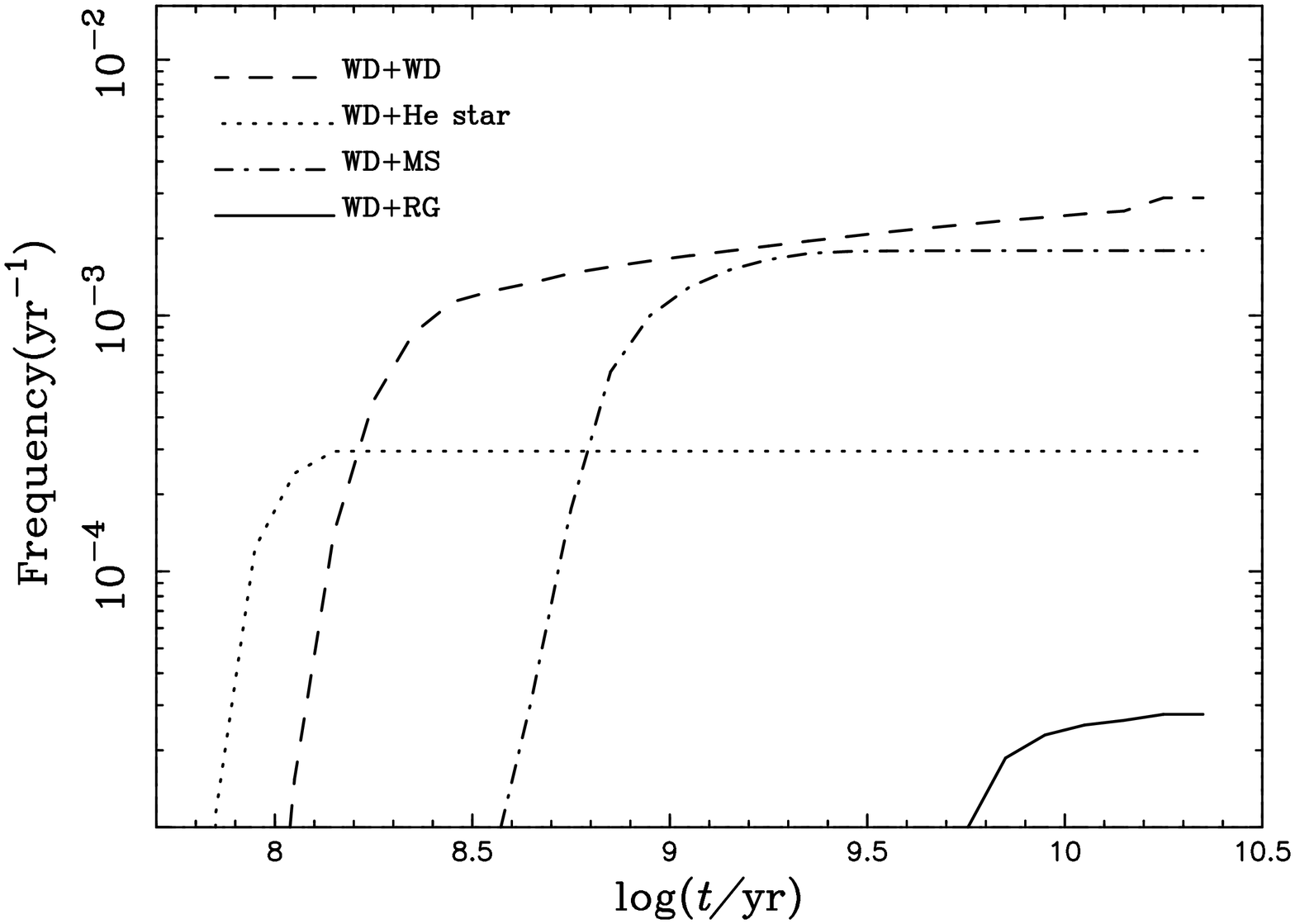}
 \caption{Evolution of Galactic SN Ia birthrates for a constant star
formation rate ($Z=0.02$, ${\rm SFR}=5\,M_{\rm \odot}{\rm
yr}^{-1}$). The key to the line-styles representing different
progenitor models is given in the upper left corner.}
\end{figure*}

\begin{figure*}
\includegraphics[width=8.5cm,angle=270]{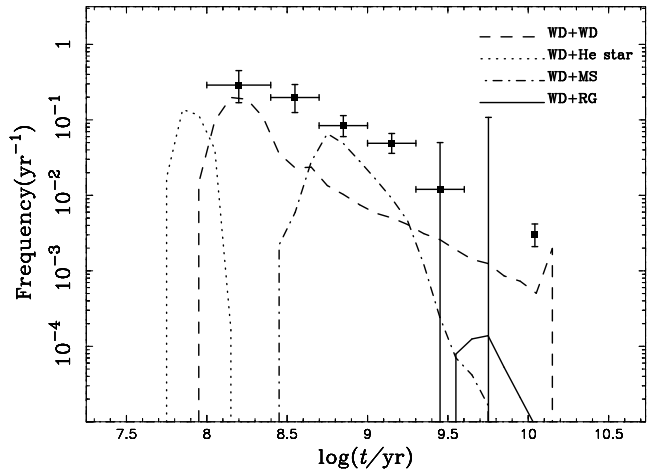}
 \caption{Similar to Figure 1,
but for a single starburst with a total mass of $10^{\rm
11}M_{\odot}$. The observational points are from Totani et al.
(2008)$^{[21]}$. }
\end{figure*}

\end{document}